\documentclass[12pt]{article}

\usepackage[final]{graphicx}
\usepackage{graphics}
\usepackage{floatflt}
\usepackage[small,bf]{caption}

\begin{document}

\title{Transverse Spin Effects at COMPASS}


\author{H. Wollny on behalf of the COMPASS collaboration\\
  \small Physikalisches Institut der Albert-Ludwigs-Universit\"at Freiburg\\ \small Hermann-Herder Str. 3, 79104 Freiburg, Germany}

\maketitle

\begin{abstract}
The measurement of transverse spin effects in semi-inclusive deep-inelastic scattering (SIDIS) is an important part of the COMPASS physics program. In the years 2002-2004 data was taken by scattering a 160\,GeV/$c$ muon beam off a transversely polarized deuteron target. In 2007, additional data was collected on a transversely polarized proton target. New preliminary results for the Collins and Sivers asymmetries from the analysis of the proton data are presented.
\end{abstract}


\section{Introduction}

Single spin asymmetries in semi-inclusive deep-inelastic
scattering (SIDIS) off transversely polarized nucleon targets have been under intense
experimental investigation over the past few years \cite{COMPASS:2005,COMPASS:2007,HERMES}.
They provide new insights into QCD and the nucleon structure. For instance, they allow the determination
of the third yet-unknown leading-twist quark distribution function $\Delta
_{T}q(x)$, the transversity distribution \cite{COLLINS_ARTRU,COLLINS}. Additionally, they give insight into
the parton transverse momentum distribution and angular momentum.

\section{The COMPASS Experiment}

COMPASS is a fixed target experiment at the CERN SPS accelerator with a wide physics program focused on the nucleon spin structure and on hadron spectroscopy. COMPASS investigates transversity and the transverse momentum structure of the nucleon in SIDIS. A 160\,GeV/$c$ muon beam is scattered off a transversely polarized hydrogen or deuterium target. The scatterd muon and the produced hadrons are detected in a wide-acceptance two-stage detector with excellent particle identification capabilities \cite{COMPASS_SPEC:2007}. In the years 2002, 2003 and 2004, data was collected on a transversely polarized $^6LiD$ target \cite{COMPASS_DEUT}. In 2007 data were taken with a transversely polarized $NH_3$ target. In the following we will focus on the new results of the data taken with the proton target.

\section{The Collins asymmetry}

The chiral-odd transversity distribution  $\Delta_T q(x)$ can be measured in combination with the chiral-odd Collins fragmentation function $\Delta_0^T D_q^h(x)$ in SIDIS.
According to Collins the fragmentation of a transversely polarized quark into an unpolarized hadron leads to a sine modulation in $\Phi_C = \phi_h + \phi_S - \pi$ in the SIDIS cross section \cite{COLLINS}. Here $\phi_h$ is the azimuthal angle of the hadron with respect to the scattering plane and $\phi_S$ is the azimuthal angle between the spin of the initial quark and the scattering plane \cite{ARTRU}. The number of produced hadrons $N$ can be written as:
\begin{equation}
N(\Phi_C) = N_0 \cdot \left ( 1 + f \cdot P_t \cdot D_{nn} \cdot A_{Coll} \cdot \sin \Phi_C \right )
\end{equation}
in which $N_0$ is the average hadron yield, $f$ the fraction of polarized material in the target, $P_t$ the target polarization and $D_{nn} = (1-y) / (1 - y + y^2/2)$ the depolarization factor.

\section{The Sivers asymmetry}

Another source of azimuthal asymmetry is related to the Sivers effect \cite{SIVERS}. The Sivers parton distribution function $\Delta_0^T q(x,{\bf k_T})$ describes the correlation of quarks with intrinsic transverse momentum ${\bf k_T}$ in a transversely polarized nucleon. The Sivers effect leads to an azimuthal modulation of the number of produced hadrons
\begin{equation}
N(\Phi_S) = N_0 \cdot \left ( 1 + f \cdot P_t \cdot A_{Siv} \cdot \sin \Phi_S \right ),
\end{equation}
with $\Phi_S = \phi_h - \phi_S$.

Since the Collins and Sivers asymmetries are othogonal to each other, both asymmetries can be determined independently from the same dataset.

\section{Data sample and event selection}

In 2007 COMPASS took data with a transversely polarized proton target ($NH_3$). The polarization of the material is $\sim 90 \%$ with a dilution factor $f$ of 0.15. The target consists of three cells in a row, where the two outer cells are polarized in one direction and the middle cell is polarized oppositely. To reduce systematics the polarization was reversed every 4-5 days. The new solenoid magnet installed in 2005 increased the angular acceptance up to $180\,$mrad.\\
The quality and the stability of the data was carefully checked. For the results presented here, about 20\,$\%$ of the collected data have been analyzed.

To select DIS events, kinematic cuts on the squared four momentum transfer $Q^2 > 1$\,(GeV/$c$)$^2$, the fractional energy transfer of the muon $0.1 < y < 0.9$ and the hadronic invariant mass $W > 5$\,GeV/$c^2$ were applied. The hadron sample consists of all charged hadrons originating from the reaction vertex with $p_T^h > 0.1$\,GeV/$c$ and $z > 0.2$. To reject muons an energy deposit of $\geq 5$\,GeV in the hadronic calorimeters was demanded.

To extract the asymmetries a maximum likelihood method binned in $\phi_h$ and $\phi_S$ was used. In each bin the product of the detector acceptance and the expression for the transverse polarization dependent part of the SIDIS cross section \cite{SIDIS_Xsection} was fitted. To disentangle acceptance and spin dependent modulations two samples of one period with opposite polarization and two samples of a consecutive period with reversed target polarization were coupled.
To fix the acceptance it was assumed that in each bin and each sample the change of acceptance between the two consecutive periods is described by one constant. The results have been checked by several other methods described in \cite{COMPASS:2005}.

\section{Results}

\begin{figure}
	  \includegraphics[width=0.93\textwidth,trim= 20 10 10 0,clip]
        	{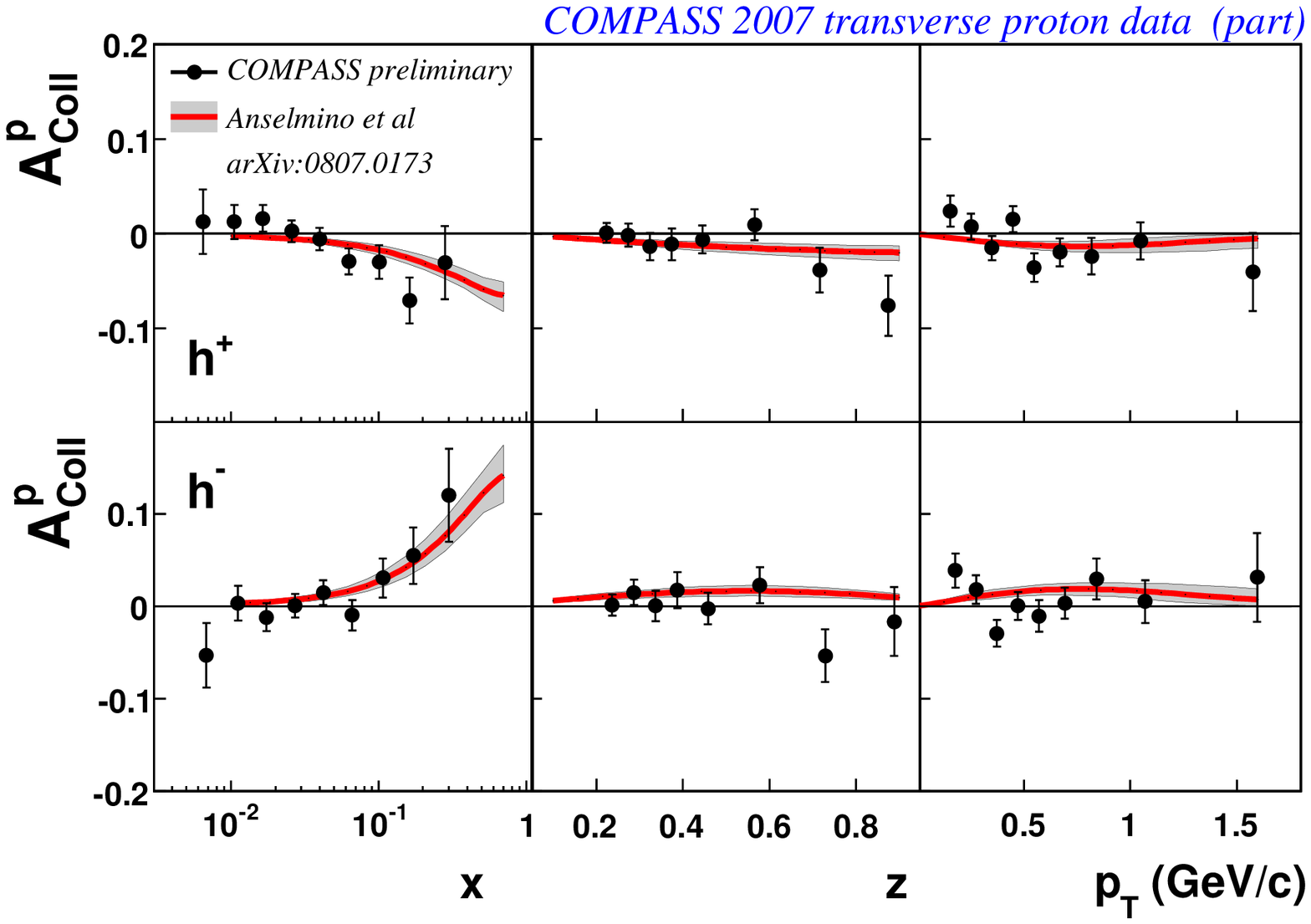} \\
	  \includegraphics[width=0.93\textwidth,trim= 20 10 10 0,clip]
        	{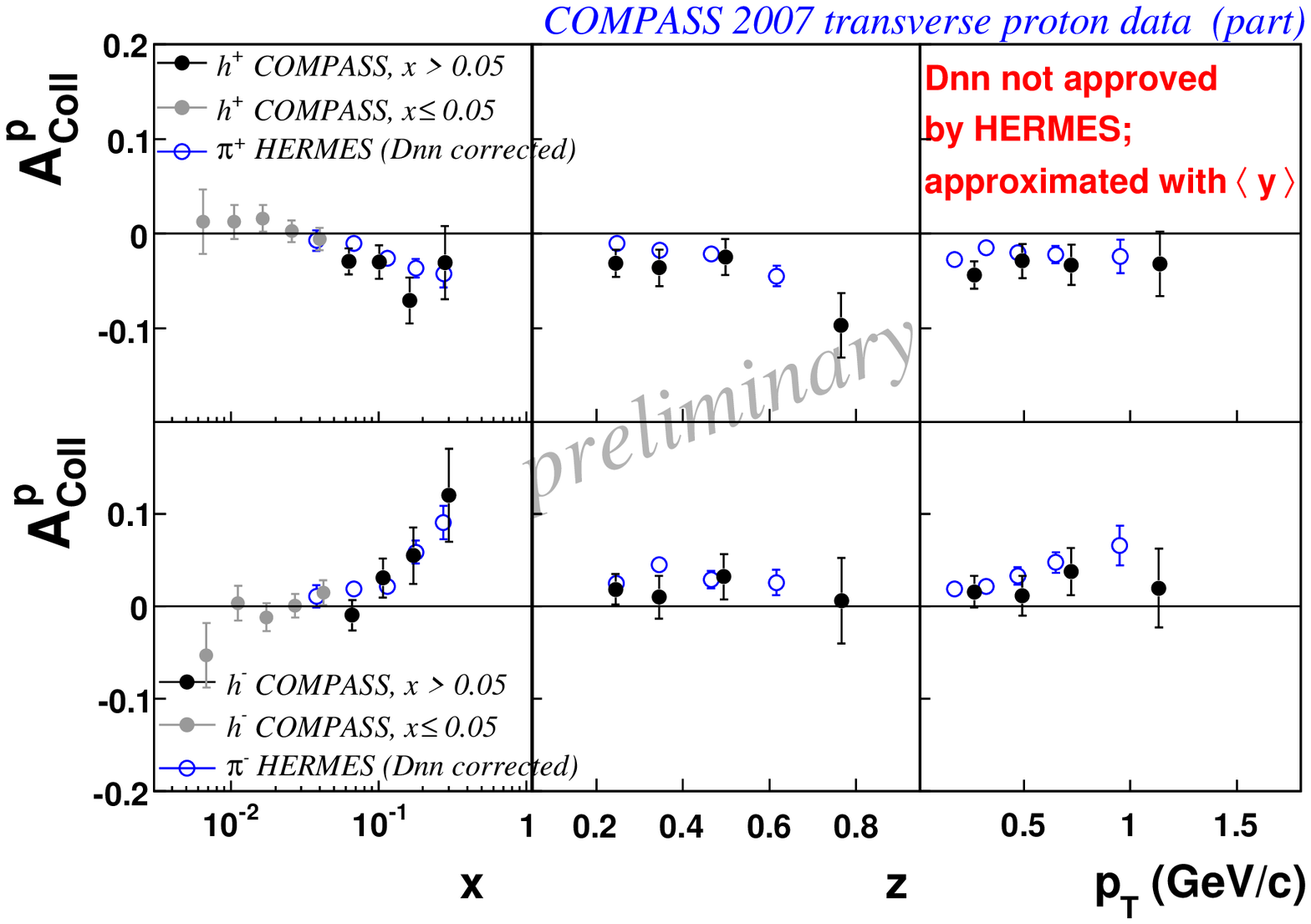}
    \caption{Collins asymmetry. Top: Comparison with recent predictions of Anselmino et al. Bottom: Comparison with results of HERMES. Only statistical errors are shown. The systematical error is 30\,$\%$ of the statistical error}
    \label{pic:collins_asym}
\end{figure}

The Collins and Sivers asymmetries were evaluated as a function of $x$, $z$, and $p_T^h$ respectively integrating over the remaining two variables. \\
In the left part of Fig. \ref{pic:collins_asym} the results for the Collins asymmetry are shown. For $x > 0.05$ the measured asymmetries are non-zero and negative in sign for positive hadrons and positive in sign for negative hadrons. For $x < 0.05$ they are small and compatible with zero. One can see the excellent agreement with the recent prediction of \cite{ANSELMINO_COLL}. The right part of Fig. \ref{pic:collins_asym} shows the comparison between COMPASS and HERMES \cite{HERMES_DIEF}. For comparing the results of the two experiments as a function of $z$ and $p_T$ a cut on $x > 0.05$ on the COMPASS data was applied and the HERMES values were corrected with $-1/D_{nn}$. The minus sign takes into account the different definitions of $\Phi_C$ and the depolarisation factor $D_{nn}(y)$ takes care of the different $y$-domains of the two experiments. For HERMES the $D_{nn}$ values for each bin have been approximated with the corresponding mean values of $y$ taken from \cite{HERMES_COLL}. The results of COMPASS and HERMES are in reasonable good agreement. 

\begin{figure}
  \includegraphics[width=0.93\textwidth,trim= 20 10 10 0,clip]
        {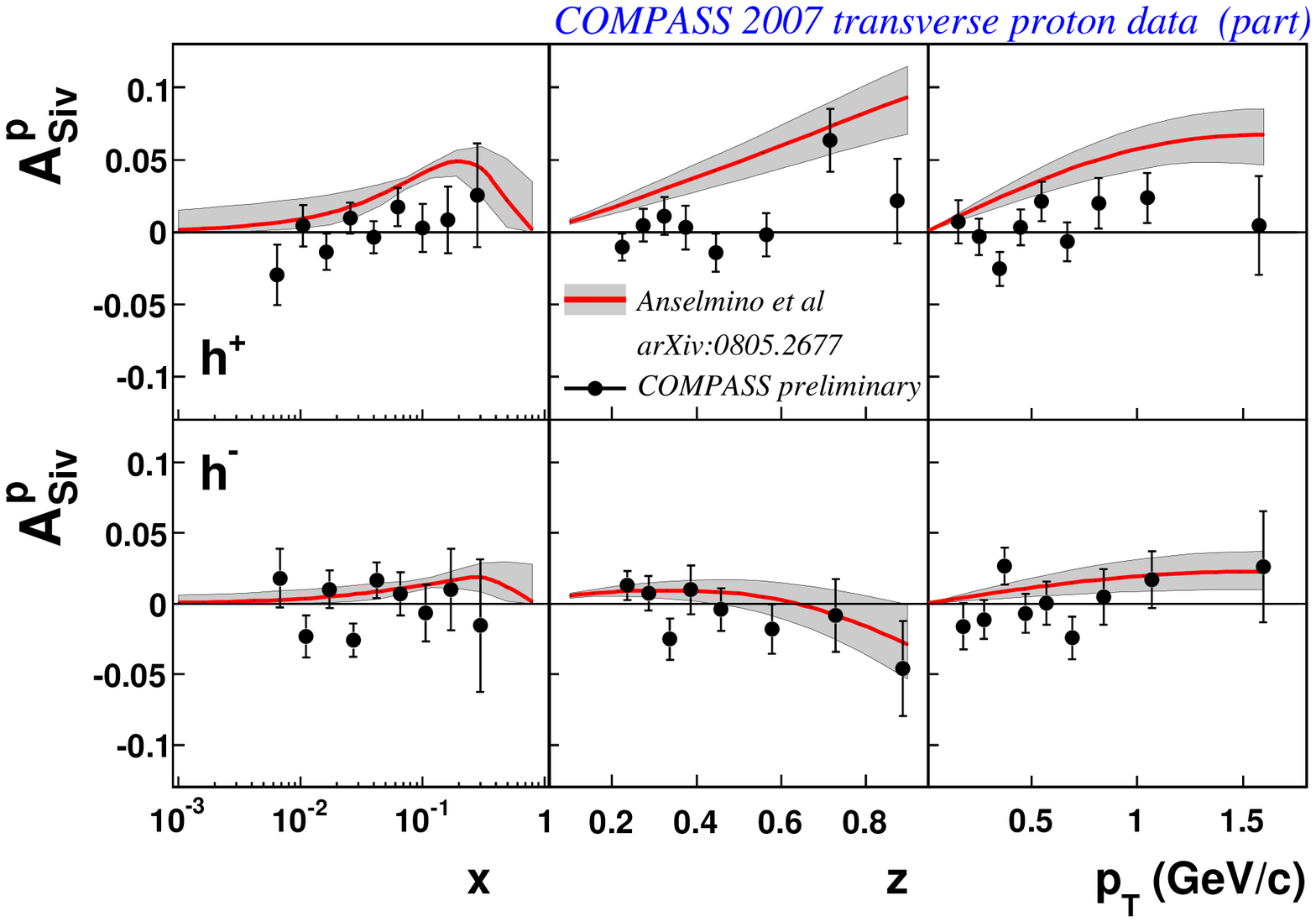} \\
  \includegraphics[width=0.93\textwidth,trim= 20 10 10 0,clip]
        {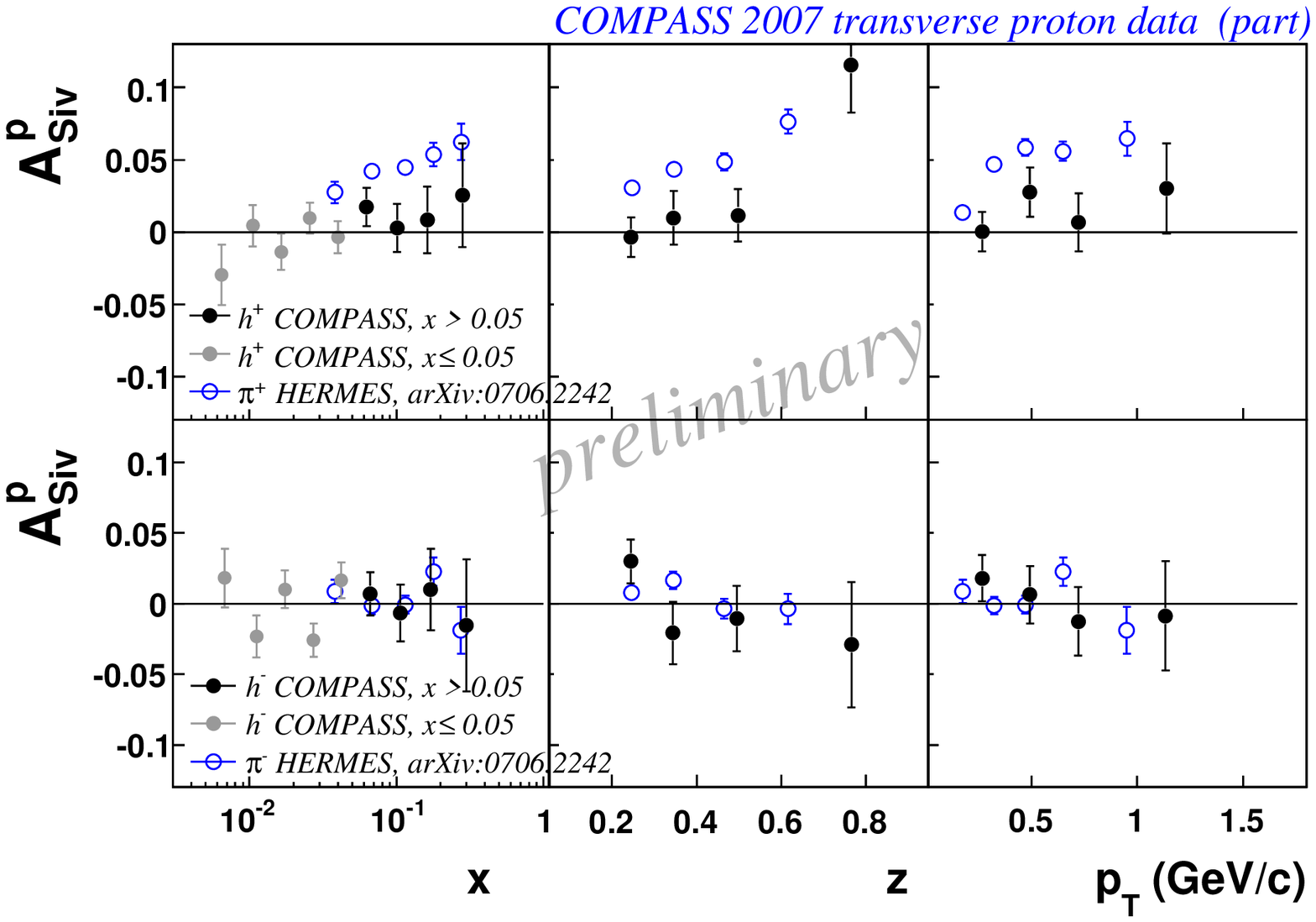}
  \caption{Sivers asymmetry.  Top: Comparison with recent predictions of Anselmino et al. Bottom: Comparison with results of HERMES. Only statistical errors are shown. The systematical error is 50\,$\%$ of the statistical error}
  \label{pic:sivers_asym}
\end{figure}

In the left part of Fig. \ref{pic:sivers_asym} the results for the Sivers asymmetry are shown. The asymmetry for both charges is small and compatible with zero within the statistical error. The comparison with the recent prediction of \cite{ANSELMINO_SIV} shows a good agreement for negative hadrons whereas it is at variance for the positive hadrons. The right part of Fig. \ref{pic:sivers_asym} shows the comparison with the results of HERMES \cite{HERMES_DIEF}. For comparing the asymmetries as a function of $z$ and $p_T$, again a cut on $x > 0.05$ was applied. The agreement for the negative hadrons is very good whereas the result for positive hadrons is at variance with HERMES. This also explains the disagreement with the prediction, since it is also based on the results of HERMES.

 \section{Summary}

First preliminary results for Collins and Sivers asymmetries measured at COMPASS in SIDIS on a transversely polarized proton target have been presented. 
The measured Collins asymmetry is non-zero for $x > 0.05$ and of opposite sign for positive and negative hadrons. For both charges the results agree very well within present statistics with recent predictions of Anselmino et al. and with the results of the HERMES group. The Sivers asymmetry for negative and positive hadrons is small and compatible with zero within the statistical precision. Whereas the result for negative hadrons is in good agreement with recent predictions of Anselmino et al. and the results of the HERMES group, the result for positive hadrons is at variance.

\bibliographystyle{unsrt}
\bibliography{wollny_heiner_spin08_proceeding}

\end{document}